# Band Anticrossing in Dilute Germanium Carbides Using Hybrid Functionals

Chad A. Stephenson,[1,a] William A. O'Brien,[1] Meng Qi,[1] Michael Penninger,[2] William F. Schneider,[2] and Mark A. Wistey[1]

[1]*Department of Electrical Engineering, University of Notre Dame, Notre Dame, Indiana 46556*
[2]*Department of Chemical and Biomolecular Engineering, University of Notre Dame, Notre Dame, Indiana 46556*

Dilute germanium carbides ($Ge_{1-x}C_x$) offer a direct bandgap for compact silicon photonics, but widely varying results on its properties have been reported. This work uses *ab initio* simulations with HSE06 hybrid density functionals and spin-orbit coupling to study the $Ge_{1-x}C_x$ band structure in the absence of defects. Contrary to Vegard's law, the conduction band minimum at $k=0$ is consistently found to decrease with increasing C content, while L and X valleys change much more slowly. A vanishing bandgap was observed for all alloys with $x>0.017$. Conduction bands deviate from a constant-potential band anticrossing model except near the center of the Brillouin zone.

## I. Introduction

Even though the size of transistors has continued to decrease, the clock speed of Si complementary metal oxide semiconductor (CMOS) chips has stagnated, while the number of CPU cores per computer is increasing exponentially. Photonic integrated circuits provide the necessary bandwidth for long distance data, but I/O, inter-core, and ultimately memory buses require considerably higher integration of all components within the logic chip itself. Si CMOS lacks an efficient, chemically-compatible laser source.

Ge and Ge alloys have received much attention due to their compatibility with Si and recent demonstrations of enhanced light emission. Liu *et al.*[1] demonstrated an optically-pumped Ge laser using small amounts of biaxial tensile strain and heavy n-type doping, with electroluminescence reported by other groups using similar techniques,[2–4] ultimately yielding an electrically-pumped laser.[5] However, the very large threshold current of the modestly strained Ge lasers and the fragility of highly strained Ge[6] make the Ge laser impractical for an efficient, integrated light emitter. GeSn has also been heavily investigated as a possible direct bandgap alloy, either as a thick metamorphic layer or grown on metamorphic InGaAs,[7–10] but device lifetimes to date have been limited.

Although both Ge and diamond emit light very weakly due to their indirect bandgap, dilute $Ge_{1-x}C_x$ alloys offer a promising route to creating lasers directly within conventional CMOS electronics. $Ge_{1-x}C_x$ is a highly-mismatched alloy; C is much more electronegative and smaller than Ge, similar to N in the GaInAsN alloy.[11,12] The N (or C) introduces an isoelectronic impurity level near the bottom of the conduction band. Due to the Pauli exclusion principle, the conduction band and impurity level repel each other, splitting the conduction band and reducing the bandgap, known as the band anticrossing (BAC) model.[11] Figure 1(a) shows a simple perturbation model demonstrating how the Γ conduction band valley decreases in energy. Although Ge is an indirect bandgap material, the direct (Γ) conduction band valley is only 140 meV above the indirect valley. Due to similar s-like (spherical) wavefunction symmetry at the Γ valley, the impurity level is expected to repel the conduction band more strongly at Γ than at L, turning $Ge_{1-x}C_x$ into a direct bandgap material.

Strong band bowing has been observed in $Ge_{1-x}C_x$ alloys with very dilute amounts of C.[13] Kolodzey *et al.* even predicted a direct bandgap alloy region with $0.04 \leq x \leq 0.11$.[14] However, others have reported linear increases in bandgap with C incorporation.[15] Differences between experimental results likely stem from defects in the material, particularly interstitial C and C-C clusters. Gall *et al.* has shown that it is much more energetically favorable for C to form nanoclusters than bond solely to Ge, in contrast to $Si_{1-x}C_x$ alloys.[16] Such defects raise doubts about parameter extraction for semi-empirical and simplified computational models.[17–19]

This work seeks to increase the accuracy of *ab initio* modeling of defect-free $Ge_{1-x}C_x$ alloys in order to extract their fundamental, intrinsic material properties, as well as to determine a target range of compositions suitable for direct bandgap devices. We use hybrid functionals with and without spin-orbit coupling (SOC) to probe the band structure at C concentrations from 0.78%-6.25%. As discussed below, the combination of small bandgap and highly-mismatched atoms invalidates many of the approximations that are typically used to reduce computational time in such simulations.

---

[a] Electronic mail: Stephenson.15@nd.edu

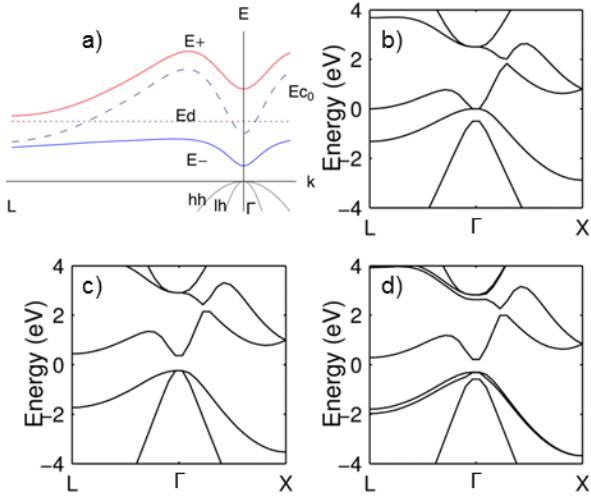

FIG. 1. a) Simple perturbation model of band anticrossing in $Ge_{1-x}C_x$. The isoelectronic impurity level $E_d$ interacts with the original conduction band $E_{c0}$ strongly at $\Gamma$, splitting the conduction band into $E^+$ and $E^-$ bands and driving $E^-$ to much lower energy at $\Gamma$, resulting in a direct bandgap. 2-atom Ge band structure simulations showing (b) inaccurate bandgap with PBE potentials, (c) opening a bandgap adding HSE06 hybrid functionals but inaccurate degenerate valence bands, and (d) improved valence bands showing accurate splitting with addition of SOC.

## II. Density Functional Theory

Density functional theory calculations were completed using the Vienna *ab initio* simulation package (VASP)[20–23], the projector-augmented wave (PAW) core treatment and a plane wave basis set with cutoff energy of 400 eV. We used the PAW PBE[24–27] potential along with HSE06 hybrid exchange-correlation functional[28]. The local density approximation (LDA) is known to underestimate the Ge bandgap to the extent that a semimetal is predicted. The generalized gradient approximation (GGA) similarly fails to reproduce the Ge band structure, Figure 1(b). We also included SOC where feasible for a more accurate representation of the valence bands and the band interactions therein.

We modeled $Ge_{1-x}C_x$ using periodic 16, 54, and 128-atom supercells in a diamond fcc lattice. These were composed of the 2-atom Ge primitive unit cell repeated 2, 3, and 4 times, respectively, along each basis vector. In each supercell, one Ge atom was replaced with a C atom. We used Gaussian smearing with Sigma=0.05 and a 9×9×9 Monkhorst-Pack k-point mesh for the 2-atom cell, which was scaled to 5×5×5, 3×3×3, and 1×1×1 for the 16-atom, 54-atom, and 128-atom supercells, respectively. In each case, the ion locations were relaxed without HSE06. Then, using HSE06, the lattice constant was varied to minimize the total system energy, i.e., to find the computational lattice constant.

Figure 1 shows the band structure of 2-atom Ge using PBE, HSE06, and HSE06 with SOC. Figure 1(d) showed reasonably good fit to Ge experimental results with a slightly underestimated bandgap of 0.59 eV at L and 0.517 eV at $\Gamma$ and spin-orbit splitting of 0.283 eV, validating our choice of HSE06 and SOC. These results could be further improved by including the semi-core d electrons as valence electrons rather than part of the core pseudopotential. However, this adds 10 more electrons per atom, and the computational cost increases 10-fold just for the 2-atom Ge calculation, and was excluded. Also, in the absence of consistent experimental reports, we did not adjust model parameters to try to fit empirical data for $Ge_{1-x}C_x$.

## III. Results
### A. 16 atom supercell

Figures 2 and 3 show band structures for $Ge_{0.9375}C_{0.0625}$ without and with spin-orbit coupling, respectively. This composition was obtained using a supercell containing 16 atoms: 15 Ge and 1 C. The computational lattice constant is 5.5744 Å, a decrease consistent with Vegard's law from the Ge computational lattice constant of 5.6953Å. With 6.25% C, the lowest conduction band is driven down in energy far enough to cross the valence bands, characteristic of conducting metals. The incorporated C has decreased the energy of the $\Gamma$ conduction band valley dramatically. This concentration of C, however, is beyond the level of a small perturbation. As is shown in the bandstructures below, the band edges have become strongly distorted in shape and are no longer approximately parabolic. Being strongly metallic, this concentration of C is also far too high for traditional, light-emitting optoelectronic materials.

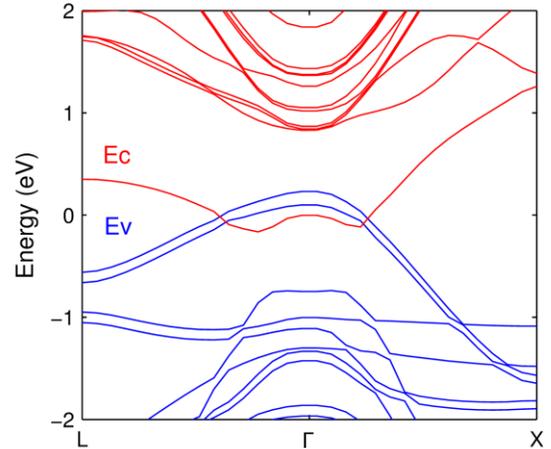

FIG. 2. Band structure of 16-atom $Ge_{1-x}C_x$ with HSE06 but without SOC.

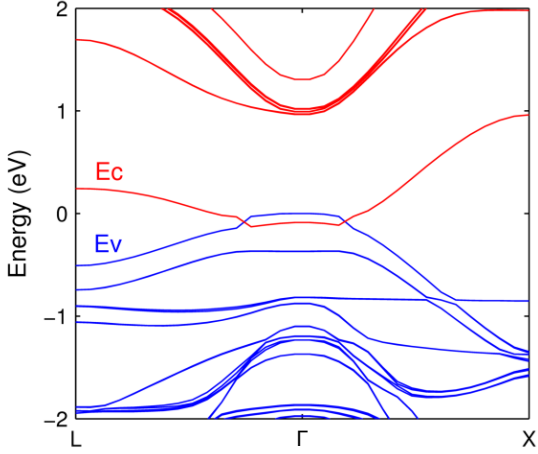

FIG. 3. Band structure of 16-atom $Ge_{1-x}C_x$ with HSE06 and SOC. Note better convergence of valence band structure with SOC.

### B. 54 atom supercell

More dilute alloys require use of larger supercells, which requires too much computation time to simulate with SOC at present. Figure 4 shows the band structure for $Ge_{0.9815}C_{0.0185}$, a supercell with 53 Ge atoms and 1 C atom. The new computational lattice constant is 5.6619 Å. For this concentration of C, the lowest conduction band at $\Gamma$ is driven to the point of just crossing the valence band. Careful inspection of the energy at the L valley shows a much smaller decrease in energy than at $\Gamma$, supporting the BAC model of strong interaction at the $\Gamma$ valley.

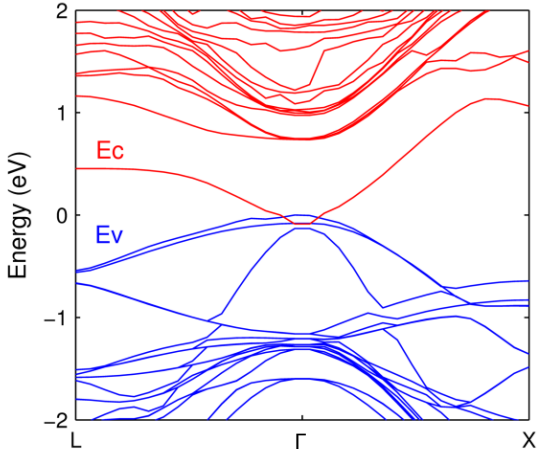

FIG. 4. Band structure of 54-atom $Ge_{1-x}C_x$ with HSE06 but without SOC. There is still no bandgap present.

### C. 128 atom supercell

Figure 5 shows the band structure of $Ge_{0.9922}C_{0.0078}$. The new lattice constant is 5.6854 Å. This figure clearly shows a strongly direct bandgap alloy, with similar conduction and valence band effective masses, and with similar shapes between the $E^+$ and $E^-$ conduction bands at $\Gamma$.

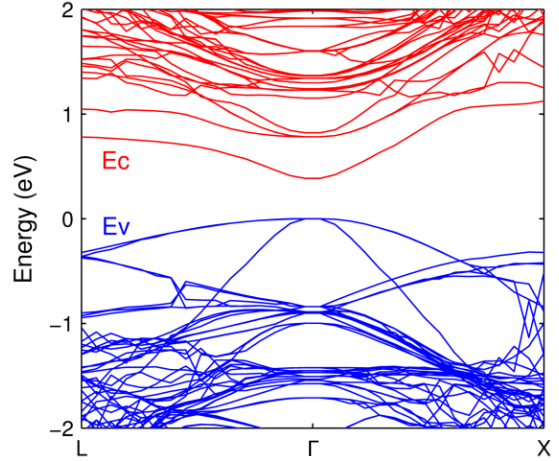

FIG. 5. Band structure of 128-atom $Ge_{1-x}C_x$ with HSE06 but without SOC. This clearly shows a direct bandgap.

### IV. Discussion

Table 1 summarizes the changes in band edge energies with percent C. The valence band maximum has been set to zero. For the 16-atom cell, the $\Gamma$ valley minimum is taken from a parabolic fit of the valley. The bandgap changes as (-170 meV ± 50)/%C for the first percent C. This decrease in the direct bandgap at $\Gamma$ is consistent with results from similar highly mismatched alloys (as much as 200 meV reduction in bandgap for 1% N in GaAs[29]) and the BAC model.

Table 1 Band edge energies for the L, $\Gamma$, and X conduction band high symmetry $k$ points with valence band maximum = 0 as reference.

| | | Band Edge Energies (eV) | | |
|---|---|---|---|---|
| %C | SOC | L | $\Gamma$ | X |
| 0.00 | yes | 0.590 | 0.517 | 0.9355 |
| 6.25 | no | 0.1187 | -0.4988 | 1.026 |
| 6.25 | yes | 0.244 | -0.1875 | 0.9605 |
| 1.85 | no | 0.4522 | -0.0886 | 1.065 |
| 0.78 | no | 0.7794 | 0.3851 | 1.123 |

In addition, the lowest L valley does not appear to change as quickly with increasing C content. Interpolation of $E_{g\Gamma}$ indicates that $Ge_{1-x}C_x$ becomes metallic ($E_g < 0$) for $x>0.017$. There is still some uncertainty in this result given the bandgap at Γ is slightly underestimated for Ge and the lack of SOC data for 54- and 128-atom supercells. Furthermore, the actual bandgap should be compared with results from $Ge_{1-x}C_x$ grown by techniques that minimize C clusters; these measurements are underway.[30]

It is noteworthy that the BAC model provides a reasonable fit to the lowest two conduction bands only near Γ or along <111> toward L. Figures 4 and 5 show the lowest CB approaching a horizontal asymptote at L, as predicted by the BAC model (see $E_d$ in Figure 1(a)). However, along <100> toward X, the CB clearly crosses this asymptote as well as the energies of the next higher conduction bands ($E^+$). This indicates that the interaction parameter $V$ in the BAC model is not constant with $k$, but varies anisotropically as a function of $\vec{k}$.

At L, two distinct bands can be identified as $E^+$ and $E^-$. This allows us to tentatively identify the energy of the carbon isoelectronic impurity level as $E_C = (E^+ + E^-)/2 = 1.06 \pm 0.19$ eV above the valence band maximum.

VASP overestimates the lattice constant for Ge by 0.66%. As we would expect due to C being a smaller atom, the calculated lattice constants of the $Ge_{1-x}C_x$ cells decrease linearly with increasing C content, as shown in Figure 6. The lattice constants follow Vegard's law, following a straight line between Ge and C, in contrast to predictions by Kelires et al.[18] The right axis shows biaxial tensile strain that would occur if grown on Ge. With 6.25% C, there is nearly 2.5% biaxial tensile strain if grown on Ge, which is unlikely to be physically realizable in bulk form. The 54-atom supercell, with 1.85% C, would have about 1% biaxial tensile strain if grown on Ge. This is still a significant strain, but it is much more realistic to achieve defect-free material. 0.78% C should be stable for bulk growth as the critical thickness is approximately 160 nm (and MBE growth can usually achieve thicker than critical thickness due to it being a far-from-equilibrium technique). One possible solution to mitigate the strain would be to incorporate dilute Sn during growth as well as C. GeSn is also expected to have a direct bandgap region, but the strain from the large Sn atom makes it especially difficult to grow high quality films.

One of the issues with tensile-strained Ge for light emission is the low occupation of electrons in the Γ valley. Even with 2% biaxial tensile strain, enough to turn Ge into a direct bandgap, only 2.5-6% of electrons are in the Γ valley.[31,32] Band anticrossing has the advantage for light emitters of increasing the effective mass in the Γ valley.[33] As the lower conduction band valley is repelled and driven down in energy at Γ, the effective mass increases. This increase leads to higher occupation at Γ, a closer match between $m_e^*$ and $m_h^*$, and therefore stronger recombination across the direct bandgap. This strongly suggests $Ge_{1-x}C_x$ as a preferred laser material over tensile-strained Ge.

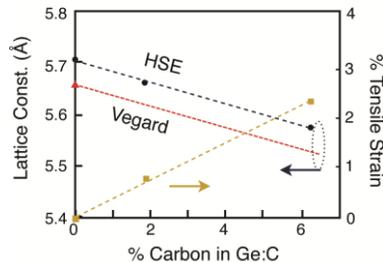

FIG. 6. Computational lattice constant compared with Vegard's law using experimental lattice constants, with increasing C content. The associated biaxial tensile strain of $Ge_{1-x}C_x$ on a Ge template is also shown.

## V. Summary

In conclusion, we computed the band structure of $Ge_{1-x}C_x$ alloys using *ab initio* hybrid exchange density functional techniques. $Ge_{1-x}C_x$ was found to be promising as a direct bandgap Group IV alloy for Si-based lasers, photodetectors, and solar cells. The band structures showed a striking reduction in $E_{gΓ}$ with increasing carbon content, estimated at (170meV ± 50)/%C for the first percent C, consistent with band anticrossing behavior at k=0. The band structure away from zone center (k>>0) was inconsistent with a constant interaction potential in the BAC model. However, a smaller change in the L valley energy suggests that the symmetry of the C isoelectronic impurity primarily affects the Γ valley, leading to a direct bandgap. Also, smaller 16-atom supercells show that spin-orbit coupling induces important changes near k=0, particularly the effective masses of electrons and light holes that are crucial for lasers and modulators. Future improvements would examine the role of biaxial strain and defects, as well as including SOC for the large supercells. These efforts are currently underway.

## Acknowledgements

This work was supported by the National Science Foundation (NSF) under grant CBET-1438608 and the Extreme Science and Engineering Discovery Environment (XSEDE) supported by NSF grant number ACI-1053575. This research was also supported in part by the Notre Dame Center for Research Computing, with valuable assistance from Dodi Heryadi. The authors thank Vince Lordi and Eoin O'Reilly for helpful discussions.